\def\comment#1{}

\newcommand{\beg}{\begin{eqnarray}}
\newcommand{\eee}{\end{eqnarray}}

\documentclass[prl,twocolumn,aps]{revtex4}
\usepackage{dcolumn}
\usepackage{amsmath2000}
\usepackage{graphicx}%
\def\cm#1{}

\begin{document}
\title{ Dual neutral variables and  
knot solitons
in triplet superconductors 
}
\author{
Egor Babaev\footnote{Email: egor@teorfys.uu.se \  http://www.teorfys.uu.se/people/egor/   
}}
\affiliation{
Institute for Theoretical Physics, Uppsala University,  
Box 803, S-75108 Uppsala, Sweden  \\
NORDITA, Blegdamsvej 17,  DK-2100 Copenhagen, Denmark 
}
\begin{abstract}
In this paper we derive a dual presentation of free energy functional
for  spin-triplet superconductors in terms of gauge invariant 
variables. The resulting equivalent model in ferromagnetic phase
has a form of a version of the Faddeev model. This allows one in particular 
to conclude that spin-triplet superconductors allow
formation of { \it stable}  finite-length closed vortices (the knotted solitons).
\end{abstract}
\maketitle
\newcommand{\la}{\label}
\newcommand{\aaa}{\frac{2 e}{\hbar c}}
\newcommand{\Pfaff}{{\rm\, Pfaff}}
\newcommand{\kA}{{\tilde A}}
\newcommand{\G}{{\cal G}}
\newcommand{\cP}{{\cal P}}
\newcommand{\M}{{\cal M}}
\newcommand{\E}{{\cal E}}
\newcommand{\btd}{{\bigtriangledown}}
\newcommand{\W}{{\cal W}}
\newcommand{\X}{{\cal X}}
\renewcommand{\O}{{\cal O}}
\renewcommand{\d}{{\rm\, d}}
\newcommand{\bfi}{{\bf i}}
\newcommand{\e}{{\rm\, e}}
\newcommand{\bfx}{{\bf \vec x}}
\newcommand{\bfn}{{ \vec{\bf  n}}}
\newcommand{\bfs}{{\vec{\bf s}}}
\newcommand{\bfE}{{\bf \vec E}}
\newcommand{\bfB}{{\bf \vec B}}
\newcommand{\bfv}{{\bf \vec v}}
\newcommand{\bfU}{{\bf \vec U}}
\newcommand{\bfp}{{\bf \vec p}}
\newcommand{\f}{\frac}
\newcommand{\bfA}{{\bf \vec A}}
\newcommand{\non}{\nonumber}
\newcommand{\be}{\begin{equation}}
\newcommand{\ee}{\end{equation}}
\newcommand{\ba}{\begin{eqnarray}}
\newcommand{\ea}{\end{eqnarray}}
\newcommand{\bastar}{\begin{eqnarray*}}
\newcommand{\eastar}{\end{eqnarray*}}
\newcommand{\half}{{1 \over 2}}
In this paper we discuss spin-triplet superconductors
and show that under certain conditions these superconductors 
should allow formation of stable finite-length solitons.
The discussions of  the 
stable finite-length solitons in 3+1 dimensions have a long history \cite{fadde}-\cite{hie}.
This concept was first discussed  in mathematical physics:
In  (\cite{fadde}) L.D. Faddeev introduced
 a  version of nonlinear $O(3)$ $\sigma$-model 
which allows formation of  the stable finite-length  solitons which 
 may have a form of a knot or a vortex loop. 
The stability of these defects in the Faddeev model is 
ensured by a  higher-order derivative term: 
\be
H_{{ F}} =  \int d{\bf r} \ \biggl[
\ \alpha  |\partial_k \bfn |^2  \ + \ \frac{1}{4e^2}
(\bfn \cdot \partial_i \bfn \times \partial_j\bfn
)^2 \biggr]
\la{fad2}
\ee
where $\bfn$ is a three-component unit vector.

Recently it was realized that versions of this model are potentially 
relevant for 
description of many different physical systems ranging 
from infrared limit of QCD to superconductivity in transition 
metals \cite{fadde}-\cite{hie}. Relevance of this 
model for condensed matter physics was pointed 
out in  Ref.  \cite{we} were
it was found that in two-band superconductors
there exists a mapping between two-flavour 
Ginzburg-Landau (GL) functional and a 
version of $O(3)$-symmetric Faddeev model. 
The knotted solitons are much more complex
and structurally complicated  topological 
defects than Abrikosov vortices and  thus 
its realization  in superconductors should open exceptionally 
wide range of possibilities of  studies of 
various phenomena 
associated with them. The remarkable circumstance is that 
the studies of the properties of these
 defects in superconductors may result 
in a ``feedback"  for the discussions of properties of 
possible similar defects in 
the infrared limit of QCD where it was argued that it describes glueballs
\cite{nature}. 

So far the two-band superconductors were the only 
known condensed matter system which is 
 described by a version of  the Faddeev model 
and allows formation of the knotted solitons
\cite{we}. 
In this paper we discuss  presentation  of the Ginzburg-Landau 
functional for a {\it single-condensate spin-triplet superconductor} 
in gauge-invariant variables by means of an {\it exact} duality mapping
which also leads to another version of the Faddeev model in ferromagnetic 
state of the condensate.
A remarkable  fact which we observe below is that the 
resulting effective  model for ferromagnetic spin-triplet superconductor 
is surprisingly  similar to the model discussed in \cite{we}
albeit  the physical origin  of this model in 
spin triplet superconductors is {\it principally} different 
from the model in \cite{we}.


We would like to emphasis that in a {\it charged} spin-1 Bose condensate
the stable knotted solitons (which 
energy is a nonmonotonic function of its size)
are the counterpart  of the Volovik-Mineev vortices
characterized by a
nontrivial Hopf invariant in  a { \it  neutral }
spin-1 superfluid
(discussed  in the pioneering papers 
 on superfluid  ${}^3 He$ \cite{vo2,vo3,vo4,vo5}).
 The  defects considered in  \cite{vo2,vo3,vo4,vo5}  have 
similar topology but its energy  is proportional to 
its size and  thus these defects are unstable against shrinkage.

We begin with reminding  a reader some important features
of the {\it neutral} spin-1 condensate following the paper
 \cite{ho}. This will be  particularly helpful in 
 the discussion of effects of coupling to gauge field and differences of the 
knotted  solitons and Volovik-Mineev vortices.

We write the order parameter of the spin-1 Bose condensate as
$\Psi_{a}({\bf r}) = \sqrt{n} ({\bf x}) \zeta_a ({\bf x})$
where  $(a=1,0,-1)$ with $\zeta$ being 
a normalized spinor $\zeta^\dagger \cdot \zeta=1$.
Then the  energy functional
for a  neutral spin-1 system can be written as \cite{ho}:
\beg
K&=& \int d {\bf r}\f{\hbar^2}{2 M} (\nabla \sqrt{n} )^2 +
 \f{\hbar^2}{2 M}n (\nabla \zeta)^2 - \mu n \nonumber \\&+&\f{n^2}{2} \left[ c_0 +c_2 <{\bf F}>^2\right], 
\label{neu}
\eee
where  $<{\bf F}> =\zeta_a^*{\bf F}_{ab}\zeta_b$ is spin.
Degenerate spinors  are related to
each other by gauge transformation $e^{i\theta}$ and spin rotations
${\cal U}(\alpha, \beta, \tau)$$=$$e^{-iF_{z}\alpha} e^{-iF_{y}\beta}
e^{-iF_{z}\tau}$,  where $(\alpha, \beta, \tau)$ are the
Euler angles.
The ground state structure of $\Psi_{a}({\bf r})$ can be found 
by minimizing the energy with fixed particle number \cite{ho}. 
Below we shall be mainly interested in  ferromagnetic state.
This state emerges when
$c_{2}<0$. The energy is minimized by 
$<{\bf F}>^2=1$ and the ground state spinor and density are \cite{ho}
\begin{eqnarray}
\zeta& =& e^{i\theta} {\cal U}
\left( \begin{array}{c} 1 \\0\\0\end{array} \right)
= e^{i(\theta-\tau)} \left( \begin{array}{c}
e^{-i\alpha}{\rm cos}^{2}\frac{\beta}{2} \\
\sqrt{2} {\rm cos}\frac{\beta}{2}{\rm sin}\frac{\beta}{2}
 \\ e^{i\alpha}{\rm sin}^{2}\frac{\beta}{2} \end{array} \right)
\nonumber \\
&&n^{o}({\bf r}) = \frac{1}{c_0+c_{2}}\mu
\label{ferrodensity} \end{eqnarray}
From this equation it is seen  that in ferromagnetic case there 
exists an  equivalence  between gauge transformation  $\theta$ and 
spin rotations $\tau$ so the 
symmetry group is  $SO(3)$ \cite{ho}.
Given the expression for the ground state spinor (\ref{ferrodensity})
one can immediately derive the superfluid velocity  for a
  neutral spin-1 ferromagnetic Bose system \cite{ho}: 
\be
{\bf v} = 
 \f{\hbar}{M}[\nabla (\theta-\tau)-\cos\beta \nabla \alpha]
\label{mh}
\ee
Let us now turn to
a {\it charged } spin-1 Bose condensate. 
We have the following expression for the free energy 
of  spin-1 superconductor in the ``ferromagnetic" state:
\beg
F&=& \int d {\bf r}\Biggl[ \f{\hbar^2}{2 M} (\nabla \sqrt{n} )^2 +
 \f{\hbar^2 n }{2 M } \left|\left(\nabla  +  i \f{2e}{\hbar c}{\bf A}\right)\zeta_a\right|^2
\nonumber \\
&-& \mu n +\f{n^2}{2} \left[ c_0 +c_2 \right] 
+\f{{\bf B}^2}{8\pi} \Biggr]
\label{cha}
\eee
with ground state spinor and density being given by (\ref{ferrodensity}).
Consequently the equation for the supercurrent is:
\beg
{\bf J} &=& \f{i \hbar e n}{M}\left( \zeta_a^* \nabla \zeta_a -\zeta_a \nabla \zeta_a^*  \right)
-\f{4 e^2 n}{M c }{\bf A}= \nonumber \\
&&  \f{ 2\hbar e n}{M}[\nabla (\theta-\tau) - \cos\beta \nabla \alpha] 
-\f{4 e^2 n}{M c }{\bf A}
\la{cur}
\eee
From this equation it is clearly seen that the supercurrent depends 
not only on phase gradients but also on spin texture what is in  direct 
analogy with the situation in neutral case (\ref{mh}). 
However, the  properties of the charged spin-1 condensate
(in particular  topological defects allowed by the system)
are principally different from the neutral case 
that can be  seen if we eliminate gauge 
field by a duality transformation to neutral variables
which  explicitly 
shows physical degrees of freedom 
in the system.  Such  a mapping, which 
is actually  different comparing to the problem  in Ref. \cite{we}, is discussed below. 

We should emphasis that 
in the charged case (\ref{cha})
the free energy  features a contribution 
from magnetic field ${\bf B}^2/8\pi$  
which can be external or self-induced  or both.
We also stress that we consider  below 
a superconductor in a simply-connected 
space, that is, our defects do not feature  zeroes 
of the order parameter. In this  case 
taking $curl$ from both sides of (\ref{cur}), (and 
taking into account that in a simply-connected space  for a regular
function holds identity $curl \nabla (\theta-\tau) =0 $)
we arrive at the following equation 
for the magnetic field in triplet superconductor 
[compare with two-gap superconductor \cite{we}]:
\beg
{ B_k} = 
-\f{ c}{4  e}[ 
\nabla_i {\cal C}_j - \nabla_j{\cal C}_i] + \f{\hbar c}{4 e} (\bfs \cdot \nabla_i
\bfs \times \nabla_j\bfs ),
\la{hhh}
\eee
where  $\nabla_i = \f{d}{dx_i}$ and we introduced the following notations:
\be
\bfs =  (\sin\beta \cos\alpha, \sin\beta\sin\alpha, \cos\beta ); 
\label{vs0}
\ee
\be 
\vec{\cal C} = \f{M}{e n}{\bf J} 
\label{vs}
\ee

Let us rearrange terms in the  Ginzburg-Landau functional  (\ref{cha}).
First let us rewrite  the second term in (\ref{cha}) as follows:
\beg
&&\f{\hbar^2 n }{2 M } \left|\left(\nabla  +  i \f{2e}{\hbar c} {\bf A}\right)
\zeta_a\right|^2  = \nonumber \\
&&
\f{\hbar^2 n }{2 M } \bigg[ \nabla \zeta_a \nabla\zeta_a^* +
   \f{2e}{\hbar c}{\bf{A} }{\bf j} + \left(\f{2e}{\hbar c}{\bf A}\right)^2 \biggr] =
\nonumber \\
&&\f{\hbar^2 n }{2 M } \bigg[ \nabla \zeta_a \nabla\zeta_a^*
 +  \left( \f{\bf j}{2} + \f{2e}{\hbar c}{\bf A}\right)^2 
- \f{{\bf j}^2}{4}\biggr] =\nonumber \\
&& \f{\hbar^2 n }{2 M } \bigg[ \nabla \zeta_a \nabla\zeta_a^*
- \f{{\bf j}^2}{4}\biggr] + \f{n}{8M}\vec{\cal C}^2
\la{11}
\eee
where 
\be
{\bf j}=\left( i\zeta_a^* \nabla \zeta_a -i\zeta_a \nabla \zeta_a^*  \right).
\ee
Then we observe the following circumstances:
\beg
\nabla \zeta_a \nabla\zeta_a^* = \f{1}{2}(\nabla \beta)^2 +(\nabla(\theta -\tau -\alpha))^2 \left[ \f{1}{2} +
\f{\cos{\beta}}{2} \right]^2 \nonumber \\
+(\nabla(\theta -\tau +\alpha))^2\left[ \f{1}{2} -
\f{\cos{\beta}}{2} \right]^2 
+(\nabla(\theta -\tau))^2\f{\sin^2\beta}{2}.
\nonumber
\eee
From the above expression it follows that:
\beg
\nabla \zeta_a \nabla\zeta_a^* - \f{{\bf j}^2}{4}= \f{1}{2}(\nabla \beta)^2 +\f{\sin^2\beta}{2}(\nabla \alpha)^2 =
\f{1}{2}(\nabla \bfs)^2
\eee
Now we can express 
the  Ginzburg-Landau functional 
for spin-1 charged ferromagnetic Bose condensate 
in the form involving only gauge-invariant variables (compare with \cite{we} ). 
That  is, with this procedure 
we eliminate the gauge field $\bf A$ and  the  variable $\theta -\tau$ 
in favor of the  gauge-invariant
variables $\bfs$ and $\vec{\cal C}$. With it the  free energy functional for
the  ferromagnetic 
$p$-wave superconductor becomes:
\beg
F&=&\int d {\bf r}\Biggl[ \f{\hbar^2}{2 M} (\nabla \sqrt{n} )^2 +
\f{\hbar^2 n}{4 M} (\nabla \bfs)^2 + \f{n}{8M}\vec{\cal C}^2
 \nonumber \\
&+&\frac{\hbar^2 c^2}{128 \pi e^2} 
\left(\f{1}{\hbar}[\nabla_i {\cal C}_j -\nabla_j 
{\cal C}_i] -\bfs \cdot \nabla_i
\bfs \times \nabla_j\bfs
\right)^2 \nonumber \\
&-& \mu n +\f{n^2}{2} \left[ c_0 +c_2 \right] \biggr]
\label{fa}
\eee 
The above expression 
is a version of the Faddeev  nonlinear $O(3)$ $\sigma$-model 
(\ref{fad2}) introduced in \cite{fadde}. The remarkable circumstance is
that this effective model for a spin-triplet superconductor 
is very similar to the effective model for two-band 
superconductors where Cooper pairs have spin-0 
but the system possesses a hidden $O(3)$-symmetry \cite{we}
and can be described by means of a three-component unit ``pseudospin"
vector.
The resulting model  explicitly displays 
 only physically relevant degrees of freedom and indeed
does not depend on $\theta$, $\tau$  and $\bf A$. 
The new variables are:  $n\equiv |\Psi|^2$, the   vector 
$\bfs$ which position on the unit sphere $S^2$
is characterized by the angles $\alpha$ and $ \beta$
(\ref{vs0}) and the massive vector field $\vec{\cal C}$. 
The effective action contains fourth-order derivative 
term for the vector field $\bfs$. 
 The field $\bfs$ 
 is also coupled to the massive vector field $\vec{\cal C}$.
Thus we can see that the properties of the charged spin-1 condensate are
very different form the neutral condensate (\ref{neu}).
Here  we would like to stress that the procedure with which 
we arrived to (\ref{fa}) is the {\it exact} change of variables 
with no approximations were used which means that there
exists the  exact mapping between the models (\ref{fa}) and (\ref{cha}).
The effective model (\ref{fa})  displays  features
of spin-triplet superconductors which can be easily overlooked 
in the presentation of the effective functional in the form (\ref{cha}).

In particular, from (\ref{fa}) we can conclude that the 
system allows {\it stable} knotted solitons characterized by a nontrivial
Hopf invariant which stability 
is ensured by the  fourth-order derivative term (the Faddeev term). 
This term is not explicitly present 
in the free energy functional in the Ginzburg-Landau form (\ref{cha}). It
stems from the term ${\bf B}^2/{8\pi}$ and has the physical meaning of 
the {\it self-induced} magnetic field in a presence of a nontrivial spin
texture (in this paper we discuss a system with no applied external fields).
 Indeed this effect  is possible due to  the feature 
of the ferromagnetic state of $p$-wave 
condensate where the magnetic field may be induced 
by a spin texture. 

 Since the effective models 
for spin-triplet superconductor and 
two-band superconductor appear being formally very similar
we refer a reader to the paper \cite{we} 
for a detailed description of the  knotted solitons 
in the model (\ref{fa}), whereas below we outline differences 
of the properties of solitons in these systems.
One of the differences is that 
 in the spin-triplet case 
the knot soliton may form as a nontrivial texture
with no inhomogeneities in Cooper pair density.
In particular that means that there is no mass 
of the components of the vector $\bfs$ whereas 
in two-band case \cite{we} the $n_3$ component 
of the unit vector $\bfn$ is related to relative 
local Cooper pair densities and is indeed  massive. 
In principle, in triplet case, the components of the 
field $\bfs$ may also acquire a mass 
if to take into account spin-orbit 
interaction which would give $\bfs$ an energetically preferred direction,
here we assume spin-orbit interaction to be small comparing 
with mass for the field $\vec{\cal C}$.
Under the assumption of small spin-orbit interaction the characteristic 
size of a knotted soliton is determined by a competition of the second-order 
and fourth-order derivative terms for $\bfs$ and thus it  is
of order of magnitude of magnetic field penetration length \footnote{
The coupling 
to the field $\vec{\cal C}$ also should affect a knot soliton: a nontrivial
configuration of texture $\bfs$ induces a magnetic field which 
in turn should induce screening Meissner current which amounts to a nontrivial 
configuration of $\vec{\cal C}$. {\it However the coupling to $\vec{\cal C}$ 
could not kill the knot soliton}. Indeed let us assume a 
knot is shrinking to a characteristic size smaller than the
magnetic field penetration length $\lambda$,
then we have a nontrivial configuration of $\bfs$
which amounts to self-induced magnetic field configuration 
proportional to $\bfs \cdot \nabla_i \bfs \times \nabla_j\bfs$
which becomes more and more singular  if  the solition is shrinking.
The direction of this self-induced field also changes inside the knot 
since the configuration of $\bfs$ is helical and is characterized 
by a Hopf invariant. On the other hand the Meissner current 
$\vec{\cal C}$ screens magnetic field in a superconductor
over a characteristic
length scale $\lambda$ thus if a texture $\bfs$
shrinks at characteristic sizes smaller than $\lambda$
we have a fast variable $\bfs$ and 
a slow variable $\vec{\cal C}$ over which 
we can average and thus at the length scales
smaller than $\lambda$ the model reduced to the model (\ref{fad2})
with an arbitrary accuracy thus the knot solitons in the 
model (\ref{fa}) are stable. This can be also 
understood intuitively: at the length scales
smaller than $\lambda$ the system almost 
does not display Meissner effect and thus 
the screening current cannot 
affect the highly nontrivial configuration of self-induced
magnetic field proportional to  $\bfs \cdot \nabla_i \bfs \times \nabla_j\bfs$.
}.
We stress that in two-band superconductor \cite{we}, in the points 
where the unit vector
$\bfn$ is situated on the south or north poles of  unit sphere $S^2$,
the densities of Cooper pair of flavour  1 or  flavour 2 vanish 
correspondingly. Since a knot soliton is characterized
by a Hopf invariant that means that the vector $\bfn$
necessary hits poles of the unit sphere which necessary
results in zeroes of Cooper pairs densities.
Also the contribution to the self-induced magnetic field associated with the term
$\bfn \cdot \partial_i \bfn \times \partial_j\bfn$ vanishes in these areas.
In contrast, the  knotted soliton  with the same topology
 in spin-triplet superconductor is a nontrivial 
configuration of spin texture and it does not feature 
zeroes of density of Cooper pairs. When spin-orbit interaction is 
small we can choose e.g. that the vector $\bfs$
 assumes at infinity the value corresponding to the 
north pole of the unit sphere and  in the knot soliton core reaches the south pole.
The configuration of the self-induced magnetic field associated with contribution 
of $ \bfs \cdot \partial_i \bfs \times \partial_j\bfs$  in spin-triplet case is following:
 the magnetic field vanishes indeed at spatial infinity (north pole of the 
unit sphere) and also the magnetic field  vanishes in the knotted soliton ``core" \footnote{ We should emphasis 
that under the {\it core}  of the  knotted soliton
 we understand here the closed line (which may form a loop or a knot)
where the vector $\bfs$ reaches the south pole. 
There are no zeros of Cooper 
pair density in what we call here the ``core". }. In between the core and ``knot boundary"
the self-induced magnetic field associated with the contribution of $ \bfs \cdot \partial_i \bfs \times \partial_j\bfs$
has a helical geometry characterized by a corresponding Hopf 
invariant. 

Thus  the knot soliton in triplet superconductor has different  
structural features and different underlying physics than the knotted solitons 
in two-gap superconductor albeit the main  feature that these defects have in common 
is the self-induced nontrivial configuration of magnetic field 
which amounts to  a Faddeev term in the effective action 
and which stabilizes the size of the soliton.
In other words 
a shrinkage of 
a knotted soliton would result in increasing energy 
of self-induced magnetic field which gives this defect 
energetic stability [compare with \cite{we,fadde,nature}]. 

Let us also  emphasis that the ``polar" phase 
of  triplet superconductors is not  described by a version 
of the Faddeev model and does  not allow formation of   knotted solitons
in contrast to the ferromagnetic phase.
The properties of a neutral polar phase  were
investigated in \cite{ho}.
This state appears in the case when $c_{2}>0$.
The energy is minimized by $<{\bf F}>=0$. The 
spinor $\zeta$ in the ground state was calculated in \cite{ho}.
In a charged counterpart of the system in the polar phase 
considered in \cite{ho}, it can be easily seen that 
the supercurrent  does not depend on the spin texture and
this  formally results in the  absence of a Faddeev term in the effective action.
 
In conclusion, we derived an exact equivalent presentation 
of the free energy functional for ferromagnetic spin-triplet superconductor. 
The derived 
functional in dual gauge-invariant variables
 explicitly displays the physical degrees 
of freedom in the system. In particular it
allows us to conclude that the ferromagnetic spin-triplet superconductors 
allow formation of stable knotted solitons. This is in contrast to a neutral
spin-1 condensate where the finite-length defects characterized 
by a nontrivial Hopf invariant are not energetically stable \cite{vo2}-\cite{vo5}.
The amazing fact is that the derived model
 is very similar to the model considered in \cite{we}
despite it was derived from a very different Ginzburg-Landau functional.
It shows  that this version of the Faddeev model 
is a rather generic model for various superconductors.
The interesting question 
is an experimental procedure how to create and observe 
the knotted solitons. 
Indeed in contrast to Abrikosov vortices these defects 
can not be created by simply applying an external magnetic field.
On the other hand  the situation is simplified by the 
 key feature of these defects, which is, once they are formed
 they are stable being protected against decay by an energy barrier.
One may expect that e.g. a rapid cooling of a system from above to
below critical temperature in e.g. applied random fields
 should indeed result in a formation of a certian density of topological defects.  
Apparently an ensemble of these defects,
 because of its complex structure, and very different nature 
comparing to  Abrikosov vortices, should exhibit many unconventional 
phenomena  and thus
should be a very interesting object for experimental studies,
especially since the triplet 
superconductivity has been 
established experimenally in several compounds 
(e.g.  $UPt_3$  and $Sr_2 Ru O_4$ \cite{machida}). 
Finally, we would like to especially stress  that 
it was suggested in \cite{nature}
that knotted solitons 
could play an important role  
in the infrared limit of QCD.
The fact   that similar  defects 
should be present in $p$-wave and two-band
superconductors and the macroscopic
quantum origin of the topological defects 
in condensed matter systems which implies
its rather direct observability means that the
spin-triplet superconductors along with 
two-band superconductors 
may in some sense serve 
as ``a testing laboratory" for
the infrared limit of QCD \cite{ppg}.

It is a great pleasure to thank  
 Prof. G. E. Volovik for 
numerous fruitful discussions and for drawing my attention to the 
triplet superconductors.
It is an equal pleasure  to thank  
Prof. L.D. Faddeev,  Prof. A.J. Niemi, Prof. Y.M. Cho
and V. Cheianov
for many useful discussions.
This work has been supported by grant
 STINT IG2001-062 and the Swedish
Royal Academy of Science.

\end{document}